\documentclass[11pt]{article}

\usepackage{amsmath}
\usepackage{amssymb}
\usepackage{latexsym}
\usepackage{graphics}
\usepackage{psfrag,fancyhdr,epsfig}

\begin{document}

\begin{titlepage}

\begin{centering}

\hfill hep-th/yymmnnn\\

\vspace{1 in}
{\bf {HIERARCHICAL NEUTRINO MASSES AND MIXING IN FLIPPED-$\bf{SU(5)}$} }\\
\vspace{1 cm}
{J. Rizos$^{1}$
 and K. Tamvakis$^{1,2}$}\\
\vskip 0.5cm
{$^1 $\it{Physics Department, University of Ioannina\\
45110 Ioannina, GREECE}}\\
\vskip 0.5cm
{$^2$\it {Physics Department, CERN, CH-1211, Geneva 23,\\
Switzerland}}

\vspace{1.5cm}
{\bf Abstract}\\
\end{centering}
\vspace{.1in}

 We consider the problem of neutrino masses and mixing in the framework of flipped $SU(5)$. The right-handed neutrino mass, generated through the operation of a seesaw mechanism by a sector of gauge singlets, leads naturally, at a subsequent level, to the standard seesaw mechanism resulting into three light neutrino states with masses of the desired phenomenological order of magnitude. In this framework
we study simple Ans{\"{a}}tze for the singlet couplings for which hierarchical neutrino masses emerge naturally as $\lambda^n\,:\,\lambda\,:\,1$ or $\lambda^n\,:\,\lambda^2\,:\,1$, parametrized in terms of the Cabbibo parameter.  The resulting neutrino mixing matrices are characterized by a hierarchical structure, in which $\theta_{13}$ is always predicted to be the smallest. Finally, we discuss a possible factorized parametrization of the neutrino mass that, in addition to Cabbibo mixing, encodes also mixing due to the singlet sector.

\vfill

\vspace{2cm}
\begin{flushleft}

December 2009
\end{flushleft}
\hrule width 6.7cm \vskip.1mm{\small \small}
 \end{titlepage}

\section{Introduction}
Despite the impressive success of the Standard Model of electroweak and strong interactions, neutrino data supply ample evidence that there is a great deal of physics beyond it. The discovery of neutrino flavour conversion establishes firmly the existence of neutrino masses and mixing. The wealth of new experimental data\cite{EXP-1}\cite{EXP-2}\cite{EXP-3}\cite{EXP-4}\cite{EXP-5} on neutrino masses and mixing angles has motivated an analogous theoretical activity aiming at uncovering the relevant mechanisms involved. A number of interesting proposals have been put forward, although the basic questions relating to the origin and structure of the neutrino mass-matrix are still standing\cite{REV}. Naturally, these attempts to understand the neutrino mass-matrix are more appealing if they are developed within the existing theoretical frameworks of grand unified theories and/or supersymmetry. Among existing proposals particularly popular is that of the so called ``seesaw mechanism"\cite{SS} giving an elegant answer to the central issue of the smallness of the neutrino mass. Apart from that, the seesaw-GUT scenario does not seem to lead by default to an understanding of the neutrino mass matrix and new ingredients are required. The quark and lepton mass matrices, although compatible with grand unification, are qualitatively different than the neutrino mass matrix. A natural explanation for this difference is provided by the seesaw mechanism in which we have a new source of mixing, not related to quarks, in the right-handed neutrino mass. Of course, these considerations vary depending on the GUT model. Overall, grand unification implies that Cabbibo mixing is expected to occur in neutrinos just as in the case of quarks. This mixing is in addition to the large mixing of neutrinos attributable to the above other source and related to physics beyond grand unification. In such a framework, while the leading part of $\theta_{12}$ and $\theta_{23}$ arises due to these effects, it is possible that the smallness of $\theta_{13}$ implies that this angle arises exclusively due to Cabbibo mixing.

Considering supersymmetric GUTs and trying to realize the seesaw mechanism, we first see that the simplest choice, namely $SU(5)$, is not so appealing, since the right-handed neutrino is a gauge singlet. As a result the right-handed neutrino Majorana mass is unconstrained and, therefore, the model lacks predictability with respect to the resulting scale of neutrino mass. In contrast, in $SO(10)$ the right-handed neutrino is part of the $\underline{16}$ spinor representation that includes all matter fermions. Nevertheless, the right-handed neutrino mass necessary for the realization of the seesaw mechanism can arise only in non-minimal versions\cite{SO(10)}. The model based on the gauge group $SU(5)\times U(1)$, the so-called {\textit{flipped}} $SU(5)$ GUT, has the interesting property that incorporates the right-handed neutrino field in the $(\underline{10},\,\underline{1})$ representation. In addition, the coupling that generates the neutrino Dirac mass is related to the up-quark Yukawa matrix. These are interesting features that present new posibilities in the realization of the seesaw-GUT scenario.

In the present article we study a supersymmetric flipped $SU(5)$ model in which an additional sector of gauge singlet superfields couple to the matter representations that contain the right-handed neutrino. As a result of these couplings the right-handed neutrino partakes in a seesaw mechanism with the singlets and obtains naturally a mass of $O(10^{12})-O(10^{15})\,GeV$. Through a subsequent standard seesaw, three light neutrino states emerge with masses of the desired order of magnitude. These masses, depending on the up-quark Yukawa coupling and the couplings to the singlet sector, are endowed with a hierarchical structure parametrised by the Cabbibo parameter as $\lambda^n\,:\,\lambda\,:\,1$ or $\lambda^n\,:\,\lambda^2\,:\,1$. Thus, for very simple singlet coupling Ans{\"{a}}tze, hierarchical neutrino masses emerge naturally. In addition, the Cabbibo-mixing part of neutrino mixing matrix comes out with the obseved hierarchical structure, in which $\theta_{13}$ is always predicted to be the smallest. We proceed further to discuss a factorized parametrization of the neutrino mass in terms of possible Ans{\"{a}}tze for the singlet couplings that encodes the dominant component of neutrino mixing, assumed to survive in the limit of vanishing Cabbibo mixing. Summarizing, two important points should be made with respect to the arising mass hierarchies, the first being that it is the extra sector of supermassive singlets and its associated seesaw mechanism that leads naturally to a desired intermediate right-handed neutrino mass scale. The second is that the derived hierarchy of light neutrino masses is related to the corresponding hierarchy of quark masses. An additional point is that the hierarchy of neutrino masses is reflected on a corresponding hierarchy of the mixing angles.

\section{The Model}
The flipped $SU(5)$ model\cite{FL} and especially its supersymmetric version\cite{FLS}, initially motivated by superstring constructions, where the adjoint representation is absent, has a number of appealing features such as the fact that neutrino masses can arise within the gauge group $SU(5)\times U(1)$, that the Higgs triplets are naturally split in mass from Higgs doublets and that baryon number-violating dimension-5 operators can be avoided. The matter (${\cal{F}},\,f^c,\,\ell^c$) and Higgs (${\cal{H}},\,{\cal{H}}^c,\,h,\,h^c$) chiral superfield content of the model is (in terms of the $SU(5)\times U(1)$  representation profile of them)
$${\cal{F}}(\underline{10},\,1)\,=\,\left({\cal{Q}},\,{\cal{D}}^c,\,{\cal{N}}^c\right),\,\,\,f^c(\underline{\overline{5}},-3)\,=\,
\left({\cal{L}},\,{\cal{U}}^c\right),\,{\cal{L}}^c(\underline{1},5),\,$$
$${\cal{H}}(\underline{10},\,1)\,=\,\left({\cal{Q}}_H,\,{\cal{D}}_H^c,\,{\cal{N}}_H^c\right),\,\,\,\,
\overline{\cal{H}}(\overline{\underline{10}},\,-1)\,=\,\left(\overline{\cal{Q}}_H,\,\overline{\cal{D}}_H^c,\,\overline{\cal{N}}_H^c\right),\,$$
$$h(\underline{5},\,-2)\,=\,\left({\cal{H}}_1,\,{\cal{D}}_H\right),\,\,\,\,h^c(\overline{\underline{5}},\,2)\,=\,\left({\cal{H}}_2,\,\overline{\cal{D}}_H\right)\,.$$
Out of these fields we may write the renormalizable cubic superpotential
\begin{equation}
{\cal{W}}_3\,=\,Y_{ij}^{(d)}\,{\cal{F}}_i{\cal{F}}_j\,h\,+\,Y_{ij}^{(u)}\,{\cal{F}}_i\,f_j^c\,h^c\,+\,Y_{ij}^{(\ell)}\,f_i^c\,{\cal{L}}_j^c\,h^c\,+\,\,\lambda{\cal{H}}{\cal{H}}\,h\,+\,
\lambda'\overline{\cal{H}}\overline{\cal{H}}\,h^c\,,\,{\label{W}}\end{equation}
where the indices are family indices. This superpotential can be augmented with a quadratic $\mu$-term ${\cal{W}}_2\,=\,\mu\,h\,h^c$. As it stands ${\cal{W}}\,=\,{\cal{W}}_3\,+\,{\cal{W}}_2$ is the most general renormalizable superpotential invariant under $R$-parity and the discrete ${\cal{Z}}_2$ symmetry that changes the sign of ${\cal{H}}\,\rightarrow\,-{\cal{H}}$, while all other fields remain unchanged. Thus, a term ${\cal{H}}{\cal{H}}^c$ cannot be present and $F$ and $D$-flatness are satisfied with the non-zero vevs
\begin{equation}\langle N_H^c\rangle\,=\,\langle\overline{N}_H^c\rangle\,\equiv\,M_X\,.\end{equation}
The fields ${\cal{Q}}_H,\,\overline{\cal{Q}}_H$ and a combination of ${\cal{N}}_H^c,\,\overline{\cal{N}}_H^c$ will be removed by the Higgs mechanism, while the triplets ${\cal{D}}^c,\,\overline{\cal{D}}^c,\,{\cal{D}}_H,\,\overline{\cal{D}}_H$ will obtain large masses $\lambda M_X,\,\lambda' M_X$ through the couplings $\lambda {\cal{H}}{\cal{H}}\,h$ and $\lambda'\overline{\cal{H}}\overline{\cal{H}}\,h^c$. Thus, the triplets are split from the doublets that remain massless. So far, the right-handed neutrino participates in a Dirac-mass term that results from
\begin{equation}
Y_{ij}^{(u)}\,{\cal{F}}_i\,f_j^c\,h^c\,\Longrightarrow\,Y_{ij}^{(u)}\,N_i^c\,\ell_j\,H_2\,+\,Y_{ij}^{(u)}\,Q_i\,u_j^c\,H_2\,+\,\dots\,.
\end{equation}

As we remarked in the introduction the large mixing encountered in neutrinos suggests that its origin is different from the corresponding Cabbibo mixing of quarks. Thus, a sector of the theory outside the GUT is required. Naturally, the fields of this sector will be {\textit{singlets}} under the GUT gauge group. The characteristic mass scale of this sector should be larger than the GUT symmetry breaking scale, presumably of the order of the string or Planck scale. Denoting these fields by ${\cal{S}}_i$, we may assign to them the $R$-parity (or matter parity) of matter superfields. Thus, the most general renormalizable superpotential that can be added to (\ref{W}) is
\begin{equation}
{\cal{W}}_S\,=\,Y^{(s)}_{ij}\,{\cal{S}}_i{\cal{F}}_j\overline{\cal{H}}\,+\,\frac{1}{2}M_{ij}^{(s)}{\cal{S}}_i{\cal{S}}_j\,.{\label{S}}
\end{equation}
For simplicity we shall restrict the number of singlet fields to just the number of generations, although a generalization to a model with more singlets is straightforward. In a generalized model with more than three singlets, the mass-term could result from a cubic term and the mass $M^{(s)}$ would be replaced by a vacuum expectation value.

\section{Neutrino Mass Scales and Hierarchies}

Thus, the superpotential of the model is the combined superpotential ({\ref{W}}) plus the $\mu$-term plus the singlet superpotential ({\ref{S}})
\begin{equation} {\cal{W}}'\,=\,{\cal{W}}\,+\,{\cal{W}}_S\,.{\label{WW}}
\end{equation}
The part of ({\ref{WW}}) that involves the neutrinos, both left and right-handed, is
\begin{equation}
Y_{ij}^{(u)}\,N_i^c\,\ell_j\,H_2\,+\,Y^{(s)}_{ij}\,S_i\,N_j^c\,\overline{H}\,+\,\frac{1}{2}M_{ij}^{(s)}S_iS_j\,.{\label{NN}}
\end{equation}
Upon symmetry breaking this will give the neutrino mass terms
\begin{equation}
Y_{ij}^{(u)}\,\frac{v_2}{\sqrt{2}}\,N_i^c\,\nu_j\,+\,Y_{ij}^{(s)}\,M_X\,S_i\,N_j^c\,+\,\frac{1}{2}M_{ij}^{(s)}S_iS_j\,,{\label{NU}}
\end{equation}
where $v_2$ is the electroweak Higgs vev $\frac{v_2}{\sqrt{2}}\,=\,\langle H_2\rangle$. Thus, neutrinos participate in the $9\times 9$ mass-matrix
\begin{equation}
\left(\begin{array}{ccc}
0\,&\,\frac{v_2}{\sqrt{2}}Y^{(u)}\,&\,0\,\\
\,&\,&\,\\
\,\frac{v_2}{\sqrt{2}}Y^{(u)}\,&\,0\,&\,Y^{(s)}M_X\,\\
\,&\,&\,\\
0\,&\,Y^{(s)}M_X\,&\,M^{(s)}\,
\end{array}\right)\,,{\label{NUMASS}}
\end{equation}
in a $\nu,\,N^c,\,S$ basis.

In the limit that the electroweak scale is neglected, the relevant part of the matrix is the $6\times 6$ matrix
\begin{equation}
\left(\begin{array}{cc}
0\,&\,Y^{(s)}M_X\,\\
\,&\,\\
\,Y^{(s)}M_X\,&\,M^{(s)}
\end{array}\right)\,.{\label{SINGLET}}
\end{equation}
As we have already remarked, the natural mass scale for the singlets should be $M^{(s)}\,>>\,M_X$. Then, it is clear that in ({\ref{SINGLET}}) a
{\textit{singlet-seesaw mechanism}} is operating that leads to the right-handed neutrino mass
\begin{equation}
M_R\,\approx\,M_X^2\,{Y^{(s)}}^{\bot}\,{M^{(s)}}^{-1}\,Y^{(s)}\,.
\end{equation}
If we take $M_X\sim\,10^{16}\,GeV$ and $M^{(S)}\sim\,10^{18}\,GeV$, for a choice of the dimensionless singlet coupling $Y^{(s)}\sim\,O(0.1)\,-O(1)\,$, we obtain the scale of $M_R$ to be $M_R\sim\,10^{12}\,-\,10^{14}\,GeV$. If we take the singlet mass scale to coincide with the string scale, we obtain $M_R\,\sim\,10^{13}\,-\,10^{15}\,GeV$.

In the limit that the three approximate mass-eigenstates with masses $O(M^{(s)})$ decouple, the neutrino mass matrix, in the $\nu,\,{N^c}'$ basis of left-handed neutrinos and ``{\textit{light}}" right-handed neutrino approximate mass-eigenstates, is
\begin{equation}
\left(\begin{array}{cc}
0\,&\,\frac{v_2}{\sqrt{2}}Y^{(u)}\,\\
\,&\,\\
\frac{v_2}{\sqrt{2}}Y^{(u)}\,&\,M_R\,
\end{array}\right)
\end{equation}
and we have the operation of the standard seesaw mechanism leading to three light neutrinos of mass
\begin{equation}
M^{(\nu)}\,\approx\,\frac{v_2^2}{2}\,Y^{(u)}\,M_R^{-1}\,Y^{(u)}\,\approx\,\frac{v_2^2}{2M_X^2}\,Y^{(u)}\,
{Y^{(s)}}^{-1}\,M^{(s)}\,(\,{Y^{(s)}}^{-1})^{\bot}\,Y^{(u)}\,.{\label{MASS}}
\end{equation}
Apart from family structure, the scale of the neutrino masses is
$$[M^{(\nu)}]\,\sim\,\left[\frac{(m^{(u)})^2}{M_R}\right]\,\Longrightarrow\,[M^{(\nu)}]_{33}\,\sim\,\frac{m_t^2}{[M_R]}\,\sim\,10^{-1}\,eV\,.$$
In the light neutrino mass formula ({\ref{MASS}}) we may factor out the mass scale
\begin{equation}
m_{\nu}\,=\,\frac{v_2^2[M^{(s)}]}{M_X^2} {\label{NUSCALE}}
\end{equation}
and replace $M^{(\nu)}\,=\,m_{\nu}\,\hat{M}^{(\nu)}$ with the dimensionless matrix
\begin{equation}
\hat{M}^{(\nu)}\,=\,Y^{(u)}\,
{Y^{(s)}}^{-1}\,\hat{M}^{(s)}\,\left(\,{Y^{(s)}}^{-1}\right)^{\bot}\,Y^{(u)}\,,{\label{NEUTRO}}
\end{equation}
where $\hat{M}^{(s)}$ is dimensionless.

It should be stressed that the right-handed neutrino mass scale was generated naturally through a seesaw mechanism in terms of the unification scale, related to the unification of gauge couplings, and the singlet sector mass scale. This would not be the case if it was introduced through a non-renormalizable term\cite{KINEZOI} the size of which has to be justified. The right-handed neutrino mass obtained this way participates in a second seesaw mechanism and gives a naturally small neutrino mass. Independently of the natural determination of neutrino scales, the formula ({\ref{MASS}}) incorporates another important feature. It combines two sources of family structure. One of them, represented by $Y^{(s)}$ and $M^{(s)}$ should endow neutrinos with the observed hard component of mixing. The other, represented by the up-quark Yukawa coupling matrix, will impart to the neutrino masses the hierarchical structure existing in the quark sector. Thus, the model, predicts naturally the {\textit{scale of neutrino masses (1)}} and, as we shall see in the remainder of this article, it has the right ingredients to provide us with {\textit{hierarchical neutrino masses (2)}} and {\textit{hierarchical neutrino mixing (3)}}.

\section{Hierarchical Neutrino Masses: Ans{\"{a}}tze}

We shall proceed now to discuss specific Ans{\"{a}}tze for the matrix structure of couplings and masses involved. We may start by adopting an Ansatz for the up-quark Yukawa matrix\cite{ROSS}, although what follows will not depend crucially on the particular choice. The essential point is that the common feature of the up-quark hierarchical mass structure will be inherited to the neutrino mass matrix. The next step is to adopt an Ansatz for the singlet coupling matrix $Y^{(s)}$. Focusing on the neutrino eigenvalues and putting aside the issue of mixing, we can proceed by adopting an Ansatz for it, that will not undo the hierarchy introduced by $Y^{(u)}$. Therefore, the singlet coupling Yukawa matrix should itself be hierarchical.
It should be noted that the Ans{\"{a}}tze of this section aim basically at obtaining the correct hierarchical structure for the neutrino mass-eigenvalues. Since, by construction, they are characterized by vanishing mixing in the limit of vanishing Cabbibo angle, they are not a priori expected to fit the mixing angle data. Nevertheless, it is interesting that certain correct features of the mixing pattern will emerge here. In particular, one of them is the {\textit{mixing angles hierarchy}}
\begin{equation}
\theta_{13}<<\theta_{12}<<\theta_{23}\,.
\end{equation}

{\textit{\textbf{Ansatz-I}}}

Let's make a specific choice of the up-quark Yukawa matrix $Y^{(u)}$ and let's adopt a simple diagonal singlet coupling matrix $Y^{(s)}$, namely
\begin{equation}
Y^{(u)}\,=\,\left(\begin{array}{ccc}
0\,&\,e_1\lambda^6\,&\,0\\
\,&\,&\,\\
e_1\lambda^6\,&\,0\,&\,e_2\lambda^2\\
\,&\,&\,\\
0\,&\,e_2\lambda^2\,&\,e_3
\end{array}\right),\,\,\,\,\,\,Y^{(s)}\,=\,{\rm Diag}\left(\,c_1\lambda^5,\,c_2\lambda^2,\,c_3\right)\,,{\label{YS-1}}
\end{equation}
where $\lambda\,\sim\,0.22$ is the Cabbibo mixing parameter. The singlet mass matrix $\hat{M}^{(s)}$ will be chosen to be a symmetric matrix with entirely generic matrix elements $\hat{M}_{ij}$.

Introducing our Ansatz into the neutrino mass formula ({\ref{NEUTRO}}), we obtain the following hierarchical eigenvalues
\begin{equation}
M_3\,\approx\,M_3^{(0)}\,+\,\lambda^2\,M_3^{(1)}\,
+\,\dots\ \ \ \ \ \ \
\end{equation}
and
\begin{equation}
M_2\,\approx\,\lambda^2\,M_2^{(0)}\,+\,\lambda^3\,M_2^{(1)}\,
+\,\dots\,,\,\,\,
\end{equation}
\begin{equation}
M_1\,\approx\,\lambda^8\,M_1^{(0)}\,+\,\lambda^9\,M_1^{(1)}\,+\,\dots\,,\,\,
\end{equation}
where $M_1^{(0)},M_1^{(1)},M_2^{(0)},M_2^{(1)},M_3^{(0)},M_3^{(1)}$ can be expressed in terms of $e_i,\,c_j$, and $\hat{M}_{ij}$.

The associated neutrino mass-diagonalization matrix is
\begin{equation}
{\bf{U}}\,=\,\left(\begin{array}{ccc}
1-\frac{b^2}{2}\lambda^6\,&\,b\lambda^3\,&\,(c-a\,b)\lambda^4\\
\,&\,&\,\\
-b\,\lambda^3\,&\,1-\frac{a^2}{2}\lambda^2\,&\,-a\,\lambda -\overline{a}\,\lambda^2\\
\,&\,&\,\\
-c\,\lambda^4\,&\,a\,\lambda +\overline{a}\,\lambda^2\,&\,1-\frac{a^2}{2}\lambda^2\\
\,&\,&\,
\end{array}\right)\,.
\end{equation}
This corresponds to a mixing matrix with
\begin{equation}
\sin\theta_{23}\,\approx\,\lambda\,a\,+\,\lambda^2\,\overline{a},\,\,\,\,\sin\theta_{12}\,\approx\,\lambda^3\,b\,,\,\,\,\,\sin\theta_{13}\,\approx\,\lambda^4\,(c-a\,b)\,
\,.\end{equation}
The coefficients $a,\,\overline{a},\,b,\,c$ are expressible in terms of the parametrers $e_i,\,c_i$ and ratios of the matrix elements $\hat{M}_{ij}$.

The adopted Ansatz has led us to the hierarchical neutrino mass-eigenvalues with approximate ratio
\begin{equation}\,\lambda^8\,:\,\lambda^2\,:\,1\,\,.\end{equation}
A hierarchy $\lambda^n\,:\,\lambda\,:\,1$ can also be obtained with a modified Ansatz. Note however that one order of magnitude in $\lambda$ can easily be overcome with an $O(1)$ numeric coefficient like $4$.

{\textit{\textbf{Ansatz-II}}}

As a second Ansatz, let us introduce the choices
\begin{equation}
Y^{(u)}\,=\,\left(\begin{array}{ccc}
0\,&\,e_1\lambda^6\,&\,0\\
\,&\,&\,\\
e_1\lambda^6\,&\,e_2\lambda^4\,&\,0\\
\,&\,&\,\\
0\,&\,0\,&\,e_3
\end{array}\right),\,\,\,\,\,Y^{(s)}\,=\,{\rm Diag}\left(\,c_1\lambda^6,\,c_2\lambda^3,\,c_3\right){\label{YU-1}}
\end{equation}
and
\begin{equation}
\hat{M}\,=\,\left(\begin{array}{ccc}
0\,&\,\hat{M}_{12}\,&\,0\\
\,&\,&\,\\
\hat{M}_{12}\,&\,\hat{M}_{22}\,&\,\hat{M}_{23}\\
\,&\,&\,\\
0\,&\,\hat{M}_{23}\,&\,\hat{M}_{33}
\end{array}\right)\,.{\label{M-II}}
\end{equation}
Note that apart from a standard choice for the up-quark Yukawa matrix\cite{ROSS} and a diagonal choice for $Y^{(s)}$ similar to ({\ref{YS-1}}), we have chosen $\hat{M}$ to possess two texture zeros.

The resulting neutrino mass-eigenvalues are
\begin{equation}
M_3\,\approx\,M_3^{(0)}\,+\,\lambda^2M_3^{(1)}\,+
\,\dots.
\end{equation}
\begin{equation}
M_2\,\approx\,\lambda M_2^{(0)}\,+\,\lambda^2M_2^{(1)}\,
+\,\dots.\,,\,
\end{equation}
\begin{equation}
M_1\,\approx\,\lambda^5\,M_1^{(0)}\,+\,\lambda^6\,M_1^{(1)}\,+\,\dots.\,.
\end{equation}
Thus, this Ansatz has led us to a neutrino mass-hierarchy
\begin{equation}
1\,:\,\lambda\,:\,\lambda^5\,\,\,\,\,\,.
\end{equation}
The diagonalizing unitary matrix is
\begin{equation}
{\bf{U}}\,=\,\left(\begin{array}{ccc}
1-\frac{a^2}{2}\,\lambda^4\,&\,a\lambda^2\,&\,-a\,b\,\lambda^3\\
\,&\,&\,\\
-a\,\lambda^2\,&\,1-\frac{b^2}{2}\,\lambda^2\,&\,b\,\lambda +c\,\lambda^2\\
\,&\,&\,\\
2\,a\,b\,\lambda^3\,&\,-b\,\lambda -c\,\lambda^2\,&\,1-\frac{b^2}{2}\lambda^2\\
\,&\,&\,
\end{array}\right)\,.{\label{XXX}}
\end{equation}
This corresponds to a mixing matrix with mixing angles
\begin{equation}
\sin\theta_{23}\,\approx\,b\,\lambda\,+\,c\,\lambda^2,\,\,\,\sin\theta_{12}\,\approx\,a\,\lambda^2,\,\,\,\,\sin\theta_{13}\,\approx\,ab\,\lambda^3\,.
\end{equation}

Although the mass patterns match the experimental values, as we anticipated earlier, neither Ansatz gives an entirely acceptable mixing pattern. For instance, $\sin\theta_{12}$ is predicted to leading order to depend only on $Y^{(u)}$ entries $\lambda\,\frac{e_1}{2e_2}$, something the excludes maximal mixing
 as $e_1,e_2$ are already fixed by the quark Yukawa couplings.

\section{Beyond Cabbibo Mixing}

This chapter is devoted to a general discussion on the issue of neutrino mixing. We begin by briefly reviewing the fundamentals. The charged lepton and neutrino mass-terms are
\begin{equation}
M_{ij}^{(\ell)}\ell_i\,\ell_j^c\,+\,M_{ij}^{(\nu)}\nu_i\nu_j\,,
\end{equation}
where $M^{(\nu)}$ is the matrix ({\ref{MASS}}). These matrices can be diagonalized as
\begin{equation}
M_{\Delta}^{(\ell)}\,=\,{{\bf{U}}^{(\ell)}}^{\bot}M^{(\ell)}{\bf{V}}^{(\ell^c)},\,\,\,\,
\,M_{\Delta}^{(\nu)}\,=\,{{\bf{U}}^{(\nu)}}^{\bot}M^{(\nu)}{\bf{U}}^{(\nu)}\,,
\end{equation}
in terms of the unitary matrices ${\bf{U}}^{(\ell)},\,{\bf{V}}^{(\ell^c)},\,{\bf{U}}^{(\nu)}$ that connect the {\textit{current}} and the {\textit{mass-eigenstates}} (primed fields)
\begin{equation}
\ell\,=\,{\bf{U}}^{(\ell)}\ell',\,\ell^c\,=\,{\bf{V}}^{(\ell^c)}{\ell^c}',\,\,\,\,\nu\,=\,{\bf{U}}^{(\nu)}\nu'\,.
\end{equation}
 The neutrino charged current
\begin{equation}
J_{\mu}^{(+)}\,\propto\,\ell_i^{\dagger}\sigma_{\mu}\nu_i\,=\,
{\ell_{\alpha}^{\dagger}}'{{\bf{U}}_{\alpha\,i}^{(\ell)}}^{\dagger}\sigma_{\mu}\,{\bf{U}}_{i\beta}^{(\nu)}\nu_{\beta}'
\,=\,{\ell_{\alpha}^{\dagger}}'\,{\cal{U}}_{\alpha\beta}\nu_{\beta}'\,
\end{equation}
can be expressed in terms of the {\textit{Pontecorvo-Maki-Nakagawa-Sakata}}\cite{PMNS} or simply {\textit{$PMNS$-mixing matrix}}
\begin{equation}
{\cal{U}}_{PMNS}\,\equiv\,{{\bf{U}}^{(\ell)}}^{\dagger}{\bf{U}}^{(\nu)}\,.{\label{PMNS}}
\end{equation}
In this paper, for reasons of simplicity, we shall not consider $CP$ violation, puting to zero all phases parametrizing $CP$. In that case, the $PMNS-$matrix can be parametrized in terms of three mixing angles, namely the {\textit{``solar angle"}} $\theta_{12}$, the {\textit{``atmospheric angle"}} $\theta_{23}$ and the {\textit{``small"}} angle $\theta_{13}$, as
\begin{equation}
{\cal{U}}_{PMNS}\,=\,\left(\begin{array}{ccc}
c_{12}c_{13}\,&\,s_{12}c_{13}\,&\,s_{13}\\
\,&\,&\,\\
-s_{12}c_{23}-c_{12}s_{23}s_{13}\,&\,c_{12}c_{23}-s_{12}s_{23}s_{13}\,&\,s_{23}c_{13}\\
\,&\,&\,\\
s_{12}s_{23}-c_{12}c_{23}s_{13}\,&\,-c_{12}s_{23}-s_{12}c_{23}s_{13}\,&\,c_{23}c_{13}
\end{array}\right)\,,
\end{equation}
where we have simplified the notation by writing $\cos\theta_{ij}=c_{ij}$ and $\sin\theta_{ij}=s_{ij}$.
Since not much is known about the charged lepton mixing matrix ${\bf{U}}^{(\ell)}$, we shall be agnostic about it, assuming only that, in general, a Cabbibo-dependent mixing matrix for the left-handed charged leptons is present. In any case, the $PMNS$ has to be equal to
\begin{equation}
{\cal{U}}_{PMNS}\,=\,{\bf{U}}(\theta_{23})\,{\bf{U}}(\theta_{13})\,{\bf{U}}(\theta_{12})\,,
\end{equation}
where the ${\bf{U}}(\theta_{ij})$ unitary matrices describe rotations in the $(i,\,j)$-plane of flavor space.

 The magnitude of neutrino mixing suggests that its major component is independent of the Cabbibo angle and originates in a sector of the theory outside the GUT. In the particular case of the model studied in this paper this source of mixing will be the singlet sector and, in particular, the couplings $Y^{(s)}$. We may parametrize this mixing through a set of {\textit{``bare"}} angles $\eta_{23}$, $\eta_{12}$. Since $\theta_{13}$ is observed to be much smaller than the other angles, a reasonable assumption would be to take a vanishing $\eta_{13}$. The overall mixing angles will have a perturbative expansion in terms of the {\textit{Cabbibo angle}} $\lambda\,\approx\,0.22$ as\cite{RAMOND}
\begin{equation}
\theta_{ij}\,=\,\eta_{ij}\,+\,\lambda\,C_{ij}^{(1)}\,+\,\lambda^2\,C_{ij}^{(2)}\,+\,\dots.{\label{ANGLES}}
\end{equation}

Going back to the neutrino mass formula,
we can always consider orthogonal transformations of the singlet couplings\footnote{The neutrino mass is invariant under rotations $Y_s'=C\,Y_s$, $M'=CMC^{\bot}$. Nevertheless, since $M$ and $M'$ are both generic and unknown, we keep the unrotated $M$ in the neutrino formula.}
\begin{equation} Y_s'\,=\,C\,Y_s\end{equation}
in terms of which it becomes
\begin{equation}
M_{(\nu)}\,=\,Y_u\,{{Y_s'}^{-1}}\,C\,M\,C^{\bot}\,{{Y_s'}^{-1}}^{\bot}\,Y_u\,.
\end{equation}
For simplicity of notation we have lowered the superscripts or dropped them altogether. We shall assume that $C$ incorporates the dominant component of neutrino mixing and that,
in the limit of vanishing Cabbibo mixing, it does not reduce to unity.
$C$ does not in general commute with the coupling matrices but we may assume that it can be chosen so that the dominant component of neutrino mixing can be factored out in terms of a general unitary matrix $U$ according to
\begin{equation}
Y_u\,{{Y_s'}^{-1}}\,C=U\,Y_u\,{{Y_s''}^{-1}}\,.{\label{ROT}}
\end{equation}
The matrix $U\,=\,{\bf{U}}(\eta_{23})\,{\bf{U}}(\eta_{12})$ describes arbitrary rotations in the $(1,2)$ and $(2,3)$ family planes and is assumed to depend only on the strong component of neutrino mixing.
The relation ({\ref{ROT}}) can be thought off as interpolating between Ans{\"{a}}tze $Y_s'$ and $Y_s''$ for the singlet coupling matrix. It should be noted that the reparametrization expressed by ({\ref{ROT}}) is not trivial since an arbitrary choice of $C$ does not always correspond to acceptable coupling matrices. Nevertheless, rewriting the neutrino mass in terms of ({\ref{ROT}}) would be rather suggestive.
Substituting these into the neutrino mass formula, we obtain
\begin{equation}
M_{(\nu)}\,={\bf{U}}(\eta_{23})\,{\bf{U}}(\eta_{12})\,\tilde{M}_{(\nu)}\,{\bf{U}}^{\dagger}(\eta_{12})\,{\bf{U}}^{\dagger}(\eta_{23})\,,
\end{equation}
where
\begin{equation}
\tilde{M}_{(\nu)}\,=\,Y_u\,{Y_s''}^{-1}M\,{{Y_s''}^{-1}}^{\bot}\,Y_u\,.{\label{TILDE}}
\end{equation}
We can now take for the quantities $Y_u,\,Y_s'',\,M$ the Ans{\"{a}}tze of the previous section (f.e. Ansatz I)
 and obtain again for $\tilde{M}_{(\nu)}$ (and $M_{(\nu)}$) the acceptable hierarchical neutrino mass eigenvalues. The mass matrix ({\ref{TILDE}})
is diagonalized by a unitary transformation
\begin{equation}
{\bf{U}}(\delta)\,=\,{\bf{U}}(\delta\theta_{13})\,{\bf{U}}(\delta\theta_{12})\,{\bf{U}}(\delta\theta_{23})\,.
\end{equation}
The mixing angles $\delta\theta_{ij}$ are proportional to powers of the Cabbibo parameter.
Thus, finally, the neutrino mass would be written in the form
\begin{equation}
M_{(\nu)}\,=\,{\bf{U}}(\eta_{23})\,{\bf{U}}(\eta_{12})\,{\bf{U}}(\delta)\,M_{(\nu)}^{(\Delta)}\,{\bf{U}}^{\dagger}(\delta)\,{\bf{U}}^{\dagger}(\eta_{12})\,{\bf{U}}^{\dagger}(\eta_{23})\,,
\end{equation}
where $M_{(\nu)}^{(\Delta)}$ is the diagonal neutrino mass matrix.
Therefore, the overall mixing matrix will be
\begin{equation}
{\bf{U}}_{PMNS}\,=\,{\bf{U}}(\eta_{23})\,{\bf{U}}(\eta_{12})\,{\bf{U}}(\delta)\,.{\label{MIX}}\end{equation}
In case a non-trivial lepton rotation matrix ${\bf{U}}_{\ell}(\lambda)$ were present, this should be canceled out by retaining an equal factor in front of the matrix $U(\eta)$ of ({\ref{ROT}}).

The mixing matrix ({\ref{MIX}}) is not yet in the desired form. We proceed by observing that for small mixing angles in $U(\delta)$
\begin{equation}
{\bf U}(\delta)=
\left(
\begin{array}{ccc}
1&\delta\theta_{12}&\delta\theta_{23}\\
-\delta\theta_{12}&1&\delta\theta_{23}\\
-\delta\theta_{13}&-\delta\theta_{23}&1
\end{array}
\right)
\end{equation}
we have
\begin{equation}
{\bf{U}}(\eta_{12}) {\bf{U}}(\eta_{23})\,{\bf{U}}(\delta)={\bf{U}}(\theta_{23}){\bf{U}}(\theta_{13}){\bf{U}}(\theta_{12})=
{\bf{U}}_{PMNS}
\end{equation}
with
\begin{eqnarray}
\theta_{12}\,&=&\,\eta_{12}\,+\,\delta\theta_{12}\\
\theta_{13}\,&=&\,\sin\eta_{12}\,\delta\theta_{23}\,+\,\cos\eta_{12}\,\delta\theta_{13}\\
\theta_{23}\,&=&\,\eta_{23}\,+\,\cos\eta_{12}\,\delta\theta_{23}\,-\sin\eta_{12}\,\delta\theta_{13}
\end{eqnarray}
For the particular case of {\textit{Ansatz-I}}, we have
\begin{equation}
\delta\theta_{23}\,\approx\,\lambda\,a\,+\,\lambda^2\,\overline{a},\,\,\,\,\delta\theta_{12}\,\approx\,\lambda^3\,b\,,\,\,\,\,\delta\theta_{13}\,\approx\,\lambda^4\,(c-a\,b)\,.
\end{equation}
It should be noted that no assumption has been made for the values of $\eta_{23}$, $\eta_{12}$, apart from the fact that a corresponding angle $\eta_{13}$ was assumed vanishing. Summarizing, in this section we have described how, a reparametrization of the singlet couplings that determine the effective right-handed neutrino mass could be set up so that the dominant part of neutrino mixing is manifest.

{\section{Conclusions.}}

In the present article we studied the problem of neutrino mass and mixing in the framework of supersymmetric flipped $SU(5)$. The right-handed neutrino field, belonging to the $(\underline{10},\,1)$ matter multiplet, was coupled to a sector of gauge singlets and obtained a mass through the operation of a seesaw mechanism. An attractive feature of the model is that the generated scale of the right-handed neutrino mass is naturally of the desired intermediate order of magnitude. This characteristic feature of the model is an immediate consequence of the fact that the mixing term to the singlets is constrained to be of the order of the GUT-breaking scale. Thus, the right-handed neutrino mass scale is a prediction of the model. At a subsequent level the model exhibits the standard see-saw mechanism resulting into three light neutrino states with masses of the desired phenomenological order of magnitude. These masses have an explicit dependence on the up-quark Yukawa coupling and the couplings to the singlet sector. Thus, they inherit a hierarchical structure parametrized by the Cabbibo parameter. In this framework
we proceeded to introduce simple Ans{\"{a}}tze for the singlet couplings for which hierarchical neutrino masses emerge naturally as $\lambda^n\,:\,\lambda\,:\,1$ or $\lambda^n\,:\,\lambda^2\,:\,1$. This is a second central property of the model. We further studied in detail the resulting neutrino mixing matrices endowed with the, thus, induced Cabbibo parameter dependence. These matrices display a hierarchical structure, in general agreement with the observed one, in which $\theta_{13}$ is always predicted to be the smallest. This is another central property of the model studied. Subsequently, we proceeded to discuss a factorized parametrization of the mixing matrix with mixing angles in the form $\theta_{ij}\,\approx\,\theta_{ij}^{(0)}\,+\,\lambda^{n(ij)}\theta_{ij}^{(1)}\,$, encoding in this way both neutrino mixing sources.

{\textbf{Acknowledgements}}

The authors wish to acknowledge the hospitality of the CERN Theory Division and the support of European Research and
Training Networks MRTPN-CT-2006 035863-1 (UniverseNet) and "UNILHC" PITN-GA-2009-237920 (Unification in the LHC era).

\end{document}